\documentclass[sigconf]{acmart}

\AtBeginDocument{}

\copyrightyear{2026}
\acmYear{2026}
\setcopyright{cc}
\setcctype{by}
\acmConference[CHIWORK '26]{Proceedings of the 5th Annual Symposium on Human-Computer Interaction for Work}{June 22--25, 2026}{Linz, Austria}
\acmBooktitle{Proceedings of the 5th Annual Symposium on Human-Computer Interaction for Work (CHIWORK '26), June 22--25, 2026, Linz, Austria}
\acmDOI{10.1145/3808045.3808061}
\acmISBN{979-8-4007-2429-9/2026/06}

\usepackage[disable]{todonotes}

\usepackage{graphicx}
\usepackage{subcaption}

\usepackage{fixltx2e}

\usepackage{enumitem}
\usepackage{array, makecell}

\def\thickhline{\noalign{\hrule height1pt}}

\newlength\replength
\newcommand\repfrac{.33}

\newcommand\rulewidth{.6pt}
\newcommand\tdashfill[1][\repfrac]{\cleaders\hbox to \replength{\smash{\rule[\arraystretch\ht\strutbox]{\repfrac\replength}{\rulewidth}}}\hfill}

\newcommand\tdotfill[1][\repfrac]{\cleaders\hbox to \replength{\smash{\raisebox{\arraystretch\dimexpr\ht\strutbox-.1ex\relax}{.}}}\hfill}

\newcommand\sub[1]{\textsubscript{#1}}

\begin{document}

\title[Barriers and Opportunities in Information Work among Blind and Sighted Colleagues]{“If We Had the Information That We Need to Interpret the World Around Us, We Wouldn’t Be Disabled:” \\ Barriers and Opportunities in Information Work among Blind and Sighted Colleagues}

\author{Yichun Zhao}
\email{yichunzhao@uvic.ca}
\affiliation{\institution{University of Victoria}
  \city{Victoria, BC}
  \country{Canada}
}

\author{Miguel A. Nacenta}
\email{nacenta@uvic.ca}
\affiliation{\institution{University of Victoria}
  \city{Victoria, BC}
  \country{Canada}
}

\author{Mahadeo A. Sukhai}
\email{mahadeo.sukhai@idea-stem.ca}
\affiliation{\institution{IDEA-STEM}
  \city{Kingston, ON} 
  \country{Canada}
}

\author{Sowmya Somanath}
\email{sowmyasomanath@uvic.ca}
\affiliation{\institution{University of Victoria}
  \city{Victoria, BC}
  \country{Canada}
}

\begin{abstract}
  Despite recognition of the value of diversity, the way work takes place can fail to support blind or low-vision employees, especially in collaborative work settings. This paper examines how professional teams with diverse visual abilities use information representations (e.g., PDF documents, spreadsheets and charts). A diary study with follow-up individual interviews (23 participants with mixed abilities from 5 teams) and 2 separate focus groups (7 participants from 2 other teams) allowed us to characterize key dimensions of the role of representations in the workplace into four types of interrelated failures and workarounds, influenced by workplace stigmas and shaped by evolving social dynamics towards interdependent information work. We contribute this new empirically supported conceptual understanding of representation use in workplaces that can help design and improve the experiences of mixed-ability teams doing knowledge work in the current technological landscape. 
\end{abstract}

\begin{CCSXML}
<ccs2012>
   <concept>
       <concept_id>10003120.10011738.10011773</concept_id>
       <concept_desc>Human-centered computing~Empirical studies in accessibility</concept_desc>
       <concept_significance>500</concept_significance>
       </concept>
   <concept>
       <concept_id>10003120.10011738.10011776</concept_id>
       <concept_desc>Human-centered computing~Accessibility systems and tools</concept_desc>
       <concept_significance>500</concept_significance>
       </concept>
   <concept>
       <concept_id>10003456.10003457.10003580.10003587</concept_id>
       <concept_desc>Social and professional topics~Assistive technologies</concept_desc>
       <concept_significance>300</concept_significance>
       </concept>
   <concept>
       <concept_id>10003456.10010927.10003616</concept_id>
       <concept_desc>Social and professional topics~People with disabilities</concept_desc>
       <concept_significance>300</concept_significance>
       </concept>
 </ccs2012>
\end{CCSXML}

\ccsdesc[500]{Social and professional topics~People with disabilities}
\ccsdesc[300]{Social and professional topics~Assistive technologies}
\ccsdesc[500]{Human-centered computing~Empirical studies in accessibility}
\ccsdesc[300]{Human-centered computing~Accessibility systems and tools}

\keywords{Knowledge work, Information work, Information accessibility, Information representation, Blind and low-vision, Mixed-visual ability team, Interdependence}

\maketitle

\section{Introduction}

Information representations are essential to how knowledge work takes place. Whether through written documents, visual diagrams, or verbal exchanges,
they allow us to externalize thoughts, organize data, and facilitate communication~\cite{Hutchins_1995, Kirsh_2010}. 
However, the act of creating and using representations can inadvertently introduce accessibility barriers, particularly for blind or low-vision (BLV) individuals. For example, consider a BLV worker who is part of a product team in an IT company, and has to access information in a PowerPoint file, which includes alt-text. Despite best efforts from the team to improve accessibility, what the alt-text and the visuals in the presentation convey are not necessarily equivalent and likely do not have similar production quality. 
There is a lack of understanding of 
how such real-world interactions related to information representations unfold for BLV workers and their sighted colleagues.

Researchers have studied mixed-ability teams operating in various domains such as in educational settings, at work, in play, and daily life~\cite{Xiao_Bandukda_Angerbauer_Lin_Bhatnagar_Sedlmair_Holloway_2024}. Studies have also explored collaborations between BLV individuals and sighted peers in contexts such as navigation~\cite{vincenzi2021interdependence}, outdoor activities~\cite{bandukda2022places}, programming~\cite{Potluri_Pandey_Begel_Barnett_Reitherman_2022}, and writing~\cite{Das_Gergle_Piper_2019, Das_Piper_Gergle_2022, Das_McHugh_Piper_Gergle_2022}. 
However, the focus in these studies has not been explicitly on information representation practices for work. 
While research like Branham and Kane's field study~\cite{Branham_Kane_2015} has provided insights into individual accessibility experiences of BLV employees within predominantly sighted 
workplaces, a gap remains in understanding the accessibility of information representations used across mixed-visual teams. 

To address this gap, we conducted a qualitative study to better understand the collective accessibility experiences of representations in-the-wild, with a focus on how real-world interactions with information representations unfold for BLV workers and their sighted colleagues. 
We explored how professional, mixed-visual settings impact access to information. Our key research questions include: 
What types of representations are used in professional work settings? 
How do these representations facilitate or hinder work and accessibility? Finally, what strategies do BLV members and others in the professional setting use to adapt and ensure work can take place effectively? 
We conducted a five-day diary study and follow-up interviews with workers from five teams, and two separate focus groups with two other teams, involving a total of 30 participants. 
This allowed us to capture rich qualitative data on the challenges, workarounds, and emergent practices, revealing a complex interplay of individual, technical, and social factors influencing information access and collaboration. 

Our paper contributes a novel characterization of information inaccessibility, focusing on four types of representational issues. We also uncover the real-world workarounds that mixed-visual ability teams engage in to negotiate those issues. 
We connect these challenges and workarounds to the top-down organizational and bottom-up social mechanisms that allow these teams to function. Finally, our findings trace the social stages through which workers progressively achieve awareness of each other's representational needs.

Despite using a mostly bottom-up representation-focused analysis, we have learned that the use of visual and spatial representations is common in mixed-ability work, that the social aspects of the work are tightly interconnected to low-level aspects of the representation, and that awareness is a base for inter-worker support that allows teams to work around the challenges and shortcomings of representations, which progressively compound with the increased complexity (and valuable functionality) of work-enabling representations.

Our study suggests designing representations with adaptive modalities to minimize negative trade-offs, while emphasizing that accessibility emerges through collective negotiation and interdependent practices involving both BLV and sighted colleagues.

 \section{Related Work}

In this section, we discuss prior work on the importance of information representations for work, approaches of making representations more accessible non-visually, and workplace practices on accessibility. 

\subsection{Information Representations for Work}
We use the term \emph{representation} to refer to \emph{external representations}, sometimes also referred to as \emph{externalizations}~\cite{zhang-NatureExternalRepresentations-1997} and \emph{knowledge representations}~\cite{Davis_Shrobe_Szolovits_1993}. Representations are objects in the work environment that encode information and allow people to interact with the information content~\cite{devries-LearningExternalRepresentations-2012, zhang-NatureExternalRepresentations-1997}. 
Larkin and Simon~\cite{larkinWhyDiagramSometimes1987} and several others~\cite{zhangRepresentationsDistributedCognitive1994,Davis_Shrobe_Szolovits_1993} emphasize that the form of a representation as well as the operations it supports are key to its effectiveness (e.g., certain visual representations allow quick navigation between related items). Therefore, we consider the full interface through which the encoding of information is accessed as part of the representation itself. For example, an Excel spreadsheet differs from its printed table version because it provides functionalities and data operations unavailable in print (e.g., real-time calculations). This is consistent with the study of representations in the subarea of Information Systems, in which system instances (e.g., a client database view) are also considered representations~\cite{burton-jonesUseEffectiveUse2013,Wand_Weber_1995}.

Using representation as a lens to analyze work might appear abstract and difficult, particularly because representations can take varied forms and serve multiple purposes (e.g., communication~\cite{Davis_Shrobe_Szolovits_1993}, problem solving~\cite{zhangRepresentationsDistributedCognitive1994} and collaboration~\cite{Hutchins_1995}). Nevertheless, there is a long tradition of this kind of analysis (e.g.,~\cite{Chalmers_1999, Suchman1995, Hutchins_1995, scaifeExternalCognitionHow1996a}), including research using a representational lens to consider the work domain specifically~\cite{Suchman1995, Clancey_1995, Jones_Chin_1998, Nielsen_2003}, such as how employees use calendars, email, and other documents to coordinate and carry out office tasks~\cite{Erickson_Danis_Kellogg_Helander_2008, Yarosh_Matthews_Zhou_Ehrlich_2013}. We chose the representational level of analysis because representations refer to each other and translate into other representations; considering only some types of representation or looking at certain types of representation in isolation could obscure important aspects of our topic of interest (mixed-ability teams). Previous researchers have made similar arguments (e.g.,~\cite{Doherty_Campos_Harrison, Ainsworth_2006, Ainsworth_2018, White_Pea_2011, Jones_Chin_1998, binksRepresentationalTransformationsUsing2022}).

\subsection{Visual Accessibility of Information Representations}

Existing research has studied the accessibility of various types of information representations and designed systems to make them individually more accessible to BLV people. 
For PDFs, which are notoriously difficult to access non-visually (e.g.,~\cite{AccessibilityCrisisinScholarlyPDFs}), research explored extracting semantic data to enable better navigation~\cite{Fayyaz_Khusro_Imranuddin_2023}. 
Data visualizations such as diagrams and charts have been converted to interactive text descriptions or narratives~\cite{Moured_Baumgarten-Egemole_Roitberg_Muller_Schwarz_Stiefelhagen_2024, Oppegaard_Xu_Hurtut_2023,Siu_SHKim_OModhrain_Follmer_2022, Zong_Lee_Lundgard_Jang_Hajas_Satyanarayan_2022}, 
tactile graphics~\cite{Khalaila_Harrison_Kim_Cashman_2025, Chen_PedrazaPineros_Satyanarayan_Zong_2025}, auditory representation using sonification techniques~\cite{Lazar_Chakraborty_Carroll_Weir_Sizemore_Henderson_2013, Chundury_Reyazuddin_Jordan_Lazar_Elmqvist_2023}, or a combination of alternative modalities and interactions~\cite{Zhang_Thompson_Shah_Agrawal_Sarikaya_Wobbrock_Cutrell_Lee_2024,Zhao_Nacenta_Sukhai_Somanath_2024,Sharif_Wang_Muongchan_Reinecke_Wobbrock_2022}. 
Multi-modal feedback mechanisms have also been developed to assist BLV users in understanding slideshow structures and creating slides~\cite{Kim_Lim_Park_2025,Zhang_Wobbrock_2023}. Spreadsheet systems have introduced auditory alerts for formula errors~\cite{Stockman_2004, Barowy_Berger_Zorn_2018} and additional views of the data to support understanding~\cite{Rahman_Bendre_Liu_Zhu_Su_Karahalios_Parameswaran_2021, Perera_Lee_Choe_Marriott_2024}. Some explored using generative AI (GenAI) to translate high-level design intent into accessible representations~\cite{Lee_Kohga_Landau_O’Modhrain_Subramonyam_2024, Huh_Peng_Pavel_2023}. 
Existing work also tries to make such innovations compatible with existing assistive technologies such as screen-readers~\cite{Zong_Lee_Lundgard_Jang_Hajas_Satyanarayan_2022} and refreshable tactile displays~\cite{RTD_Conversational}.

While these important efforts explored how to make representations accessible, they often overlook the broader social dynamics in a real-world knowledge work environment especially when collaborating with sighted colleagues. For instance, when using a shared digital whiteboard in a team meeting, sighted workers might dominate the discussion with visual references~\cite{Koutny_2024}. Even if BLV workers can access information non-visually, they are still excluded unless their alternative representations are integrated into the workflow. 
In this paper we examine how different representations are used by both BLV and sighted colleagues in real-world mixed-visual ability team settings, how they affect teamwork, and how social factors influence access.

\subsection{Workplace Practices about Visual Accessibility}

Beyond technical solutions, accessibility in the workplace is also influenced by human practices and social collaboration. Despite the desire of companies to hire and support disabled workers, BLV employees still face barriers in daily work activities~\cite{Workplace_Technology_Study} and by bureaucratic organizational policies of a profit-driven nature, which Marathe and Piper refer to as the accessibility paradox~\cite{Marathe_Piper_2025}. In response, BLV individuals often perform additional invisible work~\cite{Branham_Kane_2015} and adopt do-it-yourself practices~\cite{Cha_DIYDilemma} to make representations useful in practice. For example, BLV software developers create their own styling rules to help navigate shared digital documents~\cite{Pandey_Kameswaran_Rao_O’Modhrain_Oney_2021}. 
These show that in the absence of institutional support for accessibility, employees take the initiative to find workarounds to enable work~\cite{Mörike_Kiossis_2024}. 

The work discussed above tends to focus on one-directional experiences of how BLV professionals (as the only study participants) adapt to inaccessible work environments. In our work, we show how accessibility can also be a bi-directional process where sighted colleagues also adapt their behaviours to support access. This builds on the framework of interdependence~\cite{Bennett_Brady_Branham_2018}. Additionally, we treat information representations as central to work, and examine how team members negotiate accessibility through the use and modifications of these artefacts. 

\subsection{Relationship to Own Work}

In another paper~\cite{Zhao_Nacenta_Sukhai_Somanath_2026_CHI}, we provide an alternative analysis of the same dataset. Although the two analyses share a dataset and a methodological foundation (a representational approach to analysis), the analyses and results of the two papers are distinct.
The analysis that occupies us here is centred around the specific ways in which representations become barriers to mixed-ability work, the workarounds that people find around these barriers, and how organizations (top-down) and groups (bottom-up) address these barriers. In contrast, our other analysis focused specifically on the teams' workflow patterns when representations need to be transformed, the circumstances under which these transformations take place, and the costs of the transformation processes.

\section{Methods}

Our study involved two complementary research components: a diary study (with individual interviews) and focus group discussions. These were conducted with different groups of participants to capture both individual experiences and collective perspectives on how shared information is represented, used, and interpreted in professional teams that include both BLV and sighted team members. 

\subsection{Participants}

Participants were professional knowledge workers aged 19 years or older who used information representation tools for work in collaborative, team-based environments, with colleagues who have different visual abilities, including blind, low-vision, and fully sighted team members. 
They were recruited through multiple channels, including professional networks, disability advocacy organizations, social media platforms and community engagement efforts, to aim for a diverse representation of visual abilities and work contexts. 
We also used snowball sampling, asking participants to share the study with eligible colleagues or contacts who might be interested. 
The recruitment process began when interested individuals contacted the researchers by email. Researchers responded to confirm eligibility, answer any questions about the study, and collect contact information for a relevant workplace authority, such as a manager or supervisor. Researchers then contacted these authorities to request permission to conduct the study and asked them to distribute recruitment materials internally.

We do not report participants' professional roles to protect their anonymity due to the sensitive nature of professional contexts involving disability and risk of identifiability. 
The study included 30 participants from 7 different mixed-visual ability professional teams (T1–T7). T1–T5 participated in the diary study (including individual interviews; 14 BLV participants, 9 sighted); T6–T7 participated in focus groups (4 BLV, 3 sighted).
Team T6 works in academia, T5 is in consulting, T1 and T7 work in the legal sector, and T2-T4 are in non-profit organizations. 
Within these sectors, participants held varying levels of responsibility in a range of activities, such as data analysis, program design and advocacy, accessibility implementation, knowledge dissemination, or client deliverables. 
% The knowledge worker roles in Table~\ref{tab:p_demo} are mapped to participants based on definitions from previous literature~\cite{Reinhardt_Schmidt_Sloep_Drachsler_2011}, and participants' professional responsibilities and activities. Multiple roles are assigned to individuals who exhibit qualities of multiple roles. % The focus is on the types of knowledge work they performed, rather than their specific job titles or organizational positions, as disclosing such details would be an ethical breach of confidentiality. However, leadership roles within teams or projects can be inferred (i.e., ``controller'' indicates leadership). 
% P3\sub{TB}, P7\sub{LB}, P8\sub{LB}, P15\sub{TB} have leadership roles. 
Most participants have post-secondary education; P5\sub{LV}, P17\sub{LB}, and P30\sub{LB} completed high school. 
Table~\ref{tab:p_demo} includes other demographic information of participants. 
We use the subscript annotation ``TB'' to abbreviate ``totally blind,'' ``LB'' for ``legally blind,'' ``LV''for ``low-vision,'' and ``S'' for ``sighted'' for participant labels.

\begin{table*}[htbp]
\small
  \caption{Participants demographics. Teams T1-5 participated in the diary study (including individual interviews) and T6-7 in the focus groups. }
    \begin{tabular}{|p{1.3em}|p{2.1em}|p{1.3em}|p{4.8em}|p{6.1em}|p{4.1em}|p{17.8em}|} \thickhline
    \textbf{TID}   & \textbf{PID}   & \textbf{Age}   & \textbf{Gender} & \textbf{Visual Ability} & \textbf{Onset} & \textbf{Assistive Technologies} \\
    \thickhline
    T1    & P1\sub{TB}    & 40s   & Male   & Totally blind & Childhood & Screenreaders, Slate and stylus \\
    \hline
    T1    & P2\sub{S}    & 60s   & Female  & Sighted &       &   \\
    \thickhline
    T2    & P3\sub{TB}    & 60s   & Female  & Totally blind & 20s   & Screenreader \\
    \hline
    T2    & P4\sub{S}    & 50s   & Female 	 & Sighted &       &   \\
    \hline
    T2    & P5\sub{LV}    & 80s   & Female  &  Low vision & 60s   & Magnification, Large prints \\
    \hline
    T2    & P6\sub{S}    & 20s   & Female 	  & Sighted &       &   \\
    \thickhline
    T3    & P7\sub{LB}    & 30s   & Female   & Legally blind & Childhood & Screenreaders, High contrast, Magnification \\
    \hline
    T3    & P8\sub{LB}    & 50s   & Female   & Legally blind & Teenage & Screenreader \\
    \hline
    T3    & P9\sub{S}    & 30s   & Female   & Sighted &       &   \\
    \hline
    T3    & P10\sub{S}   & 20s   & Female   & Sighted &       &   \\
    \hline
    T3    & P11\sub{LB}   & 30s   & Female   & Legally blind & Birth & Screenreaders, Braille \\
    \hline
    T3    & P12\sub{LB}   & 50s   & Female   & Legally blind & Teenage & High contrast, Magnification, Screenreader \\
    \hline
    T3    & P13\sub{TB}   & 40s   & Male    & Totally blind & Birth & Screenreaders \\
    \hline
    T3    & P14\sub{LV}   & 40s   & Female   & Low vision & 20s   & Screenreaders \\
    \thickhline
    T4    & P15\sub{TB}   & 50s   & Female   & Totally blind & Birth & Screenreaders, Braille \\ 
    \hline
    T4    & P16\sub{TB}   & 40s   & Non-binary   & Totally blind & Birth & Screenreaders, Braille \\
    \hline
    T4    & P17\sub{LB}   & 60s   & Female  & Legally blind  & 30s   & Screenreader, Hearing aid \\
    \hline
    T4    & P18\sub{TB}   & 40s   & Female & Totally blind & Birth & Screenreaders, Slate and stylus, BrailleNote \\
    \hline
    T4    & P19\sub{S}   & 40s   & Female  & Sighted &       &   \\
    \hline
    T4    & P20\sub{S}   & 40s   & Male   & Sighted &       &   \\
    \thickhline
    T5    & P21\sub{LB}   & 40s   & Male   & Legally blind & Childhood & Screenreaders, High contrast, Magnification \\
    \hline
    T5    & P22\sub{S}   & 20s   & Female  & Sighted &       &   \\
    \hline
    T5    & P23\sub{S}   & 20s   & Male   & Sighted &       &   \\
    \thickhline
    T6    & P24\sub{LB}   & 50s   & Female  & Legally blind & 40s   & Screenreader, High contrast \\
    \hline
    T6    & P25\sub{S}   & 50s   & Female  & Sighted &       &   \\
    \hline
    T6    & P26\sub{S}   & 40s   & Male   & Sighted &       &   \\
    \hline
    T6    & P27\sub{S}   & 20s   & Female  & Sighted &       &   \\
    \thickhline
    T7    & P28\sub{LB}   & 40s   & Female  & Legally blind & 30s   & Magnification, High contrast \\
    \hline
    T7    & P29\sub{LV}   & 40s   & Female  & Low vision & 20s   & Magnification, High contrast \\
    \hline
    T7    & P30\sub{LB}   & 20s   & Male   & Legally blind & Childhood & Magnification, High contrast \\
    \thickhline
    \end{tabular}\label{tab:p_demo}\end{table*}

\subsection{Data Collection}

We conducted a diary study (with individual interviews) over five working days during normal working hours, involving participant teams T1-T5, to capture the in-the-moment information practices. 
To ensure data quality and consistency, participants received a pre-diary briefing that explained key concepts and the study procedure, such as what constitutes shared information and how to describe its representation and use. 
During the diary collection phase, participants recorded their interactions with shared information representations in their daily work through text or voice entries, supplemented by attached representations, screenshots of representations, or detailed descriptions. 
Each entry detailed what the representation is (e.g., a table, flowchart, or a database view), its purpose (e.g., financial budgeting), collaborators involved, and participants' activities working with the representation. 
To support participants and ensure consistent diary writing, researchers conducted daily check-ins via email or quick call throughout the study period. 
After the diary period, each participant took part in a remote post-diary interview (1-1.5 hours) to reflect on their entries and provide deeper insights into their experiences. Topics included factors influencing the choice of specific representations, impact of visual ability on the use of representation and working with colleagues, and reflections on why accessibility barriers happened and what facilitated effective teamwork using representations. We collected demographic data at the beginning of the interview.

We also conducted focus groups with teams T6 and T7 who were interested in participating but do not have the capacity for the diary study, offering an inclusive participation option. This format provided an alternative method to help uncover collective experiences on how information representations are used beyond individual accessibility considerations (though the diary and interview data also include them). 
Prior to the focus groups, participants were asked to reflect on their experiences with information representation use and complete a demographics survey. 
Each team participated in one focus group session, which lasted one-hour and was conducted remotely. 
The beginning of each session was for participants to share concrete examples of how information representations are used in their teams, creating a shared context before specific questions. 
The discussion topics are the same as those explored during post-diary interviews (described above).

Both the diary study and focus groups emphasized participant agency in describing their experiences rather than prescribing solutions, aligning with the study's exploratory focus. 
We collected 404 diary entries (including 209 examples of information representations), 29 hours and 42 minutes of interview recordings, 
and 2 hours of focus group recordings. All recordings were transcribed.
Participants received \$50 each for the diary study, or \$20 for the focus group. All study procedures were approved by our institution's ethics board.

\subsection{Data Analysis} 

We used thematic analysis~\cite{Braun_Clarke_2023}.  
The first author started by analyzing the diaries from two randomly selected teams of participants (T1 and T3) and performed open-coding, identifying 
early codes centered on participants' practices working with information representation, such as ``difficult to verify status of editing'' and ``using shortcut keys to navigate semantic structure.'' 
Next, the first author open-coded the two teams' interview transcripts, and new codes uncovered the social and workplace-related dimensions, such as ``advocating to educate about information accessibility'' and ``conflicting workplace priorities.'' 
During this stage, co-authors met periodically to iteratively refine codes, and discussed initial categorization of the codes such as workplace factors in representation accessibility and impact of representation access practices. As new codes were formed, earlier data were revisited to apply the new codes. Over time, we reached consensus on an initial codebook. 
The first author then continued coding the remaining diaries, interviews, and focus group transcripts. The research team continued meeting frequently as analysis progressed to refine the codebook and construct initial themes. The final codebook\footnote{Included in supplementary material.} consists of 80 codes.
We then collectively reviewed themes cutting through the various categories of codes, refined them, and found new ones. 
For example, the challenges of ``structural information not semantically labelled'' and ``screen magnification loses visual context'' led to the practices of ``removing structure for linear access'' and ``adding semantic labels for structural access,'' which resulted in ``reduced productivity and accuracy'' and ``lost computational features.'' These relationships between codes and categories led to the theme ``Non-Visual Access to Visual Structure.'' 
At the end, the analysis resulted in three high-level themes and eight detailed themes (Table~\ref{tab:themes}). We also invited participants to review our analysis results with illustrative quotes to avoid exaggerating potential problem areas and opportunities for improvement~\cite{McKim_2023, Smyth_Kumar_Medhi_Toyama_2010}, and five responded and validated them.

\begin{table*}[htbp]

\centering
\caption{High-level themes and detailed themes of our analysis. They map to the headings and subheadings of the Results section.}
\begin{tabular}{l l}
\toprule
\textbf{High-Level Theme} & \textbf{Detailed Theme} \\

\midrule
Representational Failures and Workarounds & Not Representing Content Non-Visually \\
 & Non-Visual Access to Visual Structure \\
 & System Feedback When Editing \\
 & Cognitive Support \\

\midrule

Accessibility As a Work Priority & \\ 

\midrule

Social Adaptations for Representational Accessibility & Advocating for Representational Accessibility \\ 
 & Assuming Representational Access Needs \\
 & Inquiring about Representational Access Needs \\
 & Building Interdependence for Representational Access \\

\bottomrule
\end{tabular}
\label{tab:themes}
\end{table*}

\subsection{Positionality Statement}

The research team comprises authors with diverse expertise regarding accessibility and information work. One author brings direct lived experience as a BLV working professional, while others have background in accessibility-related employment, teaching and research from four to seven years. One also possesses extensive experience in representation design. 
These positionalities directly affected data analysis and interpretation. 
For instance, %during our collaborative data analysis discussions, 
the researcher with information representation design expertise focused on understanding the phenomena through the technical properties of artefacts, while the BLV and accessibility researchers interpreted barriers through the lens of social stigma in workplace interactions. 
Together, these perspectives allowed us to characterize interdependent information work, capturing both the technical failures of representations and the social dynamics required to overcome them. We recognize that there are multiple models of disability (e.g., medical~\cite{Marks_1997} and social~\cite{Shakespeare_1997}) which provide differing perspectives for understanding accessibility. Our analysis is primarily informed and aligned with a relational model of disability~\cite{Shakespeare_2006, Thomas01012004, Garland-Thomson_2014, Garland-Thomson_2011}, which emphasizes the relationships between individuals, technologies, and social environments.

 \section{Results}

Participants in our study engaged with a wide range of representations in their work, summarized in Table~\ref{tab:reps}.
This includes a variety of representation types that are sometimes eminently spatial (e.g. spreadsheets, diagrams, slideshows) or even mostly visual (slideshows, some PDFs). This validates the importance of our main goal of understanding mixed-ability representational environments; our data is a clear indication that teams where visual ability is not homogeneous across team members cannot, or do not want to, completely avoid visual or spatial formats for key representations. This is despite of the obvious and less obvious barriers and challenges that many of these representational systems pose to accessibility (the main focus of Subsection~\ref{repproblems} below), the workplace rules and priorities (Subsection~\ref{workpriority}) and the social adaptations that it requires (Subsection~\ref{socialadaptation}). 
In fact, BLV team members explicitly acknowledged the role and need for such tools. For example, P8\sub{LB} explained that their team relied on financial tracking forms and visualizations which \textit{“are full of information that [the managers] have to very carefully review.”} However, despite their informational value, these representations were not meaningfully accessible. P8\sub{LB} noted that while \textit{“there’re components that I can navigate with JAWS [screenreader], but then it stops. It doesn’t go all the way [rendering it] completely useless.''}
This indicates that although such representations are necessary for the teams’ work, their limited accessibility makes them inadequate for BLV users.

We structure our findings in three parts: 1) the types of failures of representational inaccessibility that we observed, together with their workarounds, 2) the organizational expectations and rules that are affected by information accessibility, and 3) the progressive processes that enable mixed-ability teams to learn to work together.

\begin{table*}[h]
\centering
\caption{Summary of Information Representations Used.}
\begin{tabular}{l c p{9.5cm}}
\toprule
\textbf{Representation} & \textbf{Counts} & \textbf{Function} \\
\midrule
Spreadsheets & 22 & Real-time data updates, data tracking, and creating reports or visualizations \\
Databases & 10 & (As above) \\
Slideshows & 17 & To illustrate concepts and as persistent artefacts in meetings \\
Diagrams or charts & 9 & (As above) \\
Videos & 4 & (As above) \\ 
Word processing documents & 73 & Detailed documentation and information sharing \\
PDF documents & 16 & (As above) \\
Emails & 91 & Coordination and short-term exchanges \\
Text messages & 30 & (As above) \\
\bottomrule
\end{tabular}
\label{tab:reps}
\end{table*}

\subsection{Representational Failures and Workarounds}\label{repproblems}
Representations, especially visual and spatial ones, have been previously shown to present many barriers and challenges for BLV knowledge workers (e.g.,~\cite{Jung_Mehta_Kulkarni_Zhao_Kim_2022}). Nevertheless, it is important for us to, not only confirm previous findings, but also complement them. We do this in the form of a classification that exposes four distinct ways in which representations are inadequate or present significant challenges: invisibility of visual content, lack of access to visual structure, inadequate system feedback and feedthrough, and lost cognitive support. For each of these types, we explain the main representational pitfall and the workarounds and solutions that our participants found or proposed.

\subsubsection{Not Representing Content Non-Visually}

Participants reported multiple examples of representations that omitted consideration of different visual abilities, making them completely or almost completely imperceptible to BLV workers (\textit{n}=7). 
P1\sub{TB} exemplified this: \textit{``[For] paper print-out with the list of [client information ...] I could not read the list.''} P3\sub{TB}, P6\sub{S} and P17\sub{LB} documented instances where colleagues were unaware of email attachments. 
Even long-existing software features can fall into this category: 
P7\sub{LB} described how Word's correction tracker relies on visual cues like strikethroughs and highlights which, due to the lack of equivalent screen-reader feedback, leave edit history unrepresented to blind users. 
Furthermore, P13\sub{TB} shared his screen and demonstrated in his interview how a pop-up in a database view obscured essential underlying information, and his screen-reader could not detect the obstructed content. 

\paragraph{Workaround: None}
When information is not represented perceptibly, it results in total exclusion from individual information access, leaving no workarounds besides preventing the exclusion from occurring in the first place. 

\subsubsection{Non-Visual Access to Visual Structure}
\label{subsubsec:structure}

\begin{figure*}[h!]
    \centering
    \begin{subfigure}[b]{0.355\textwidth}
        \centering
        \includegraphics[width=\textwidth]{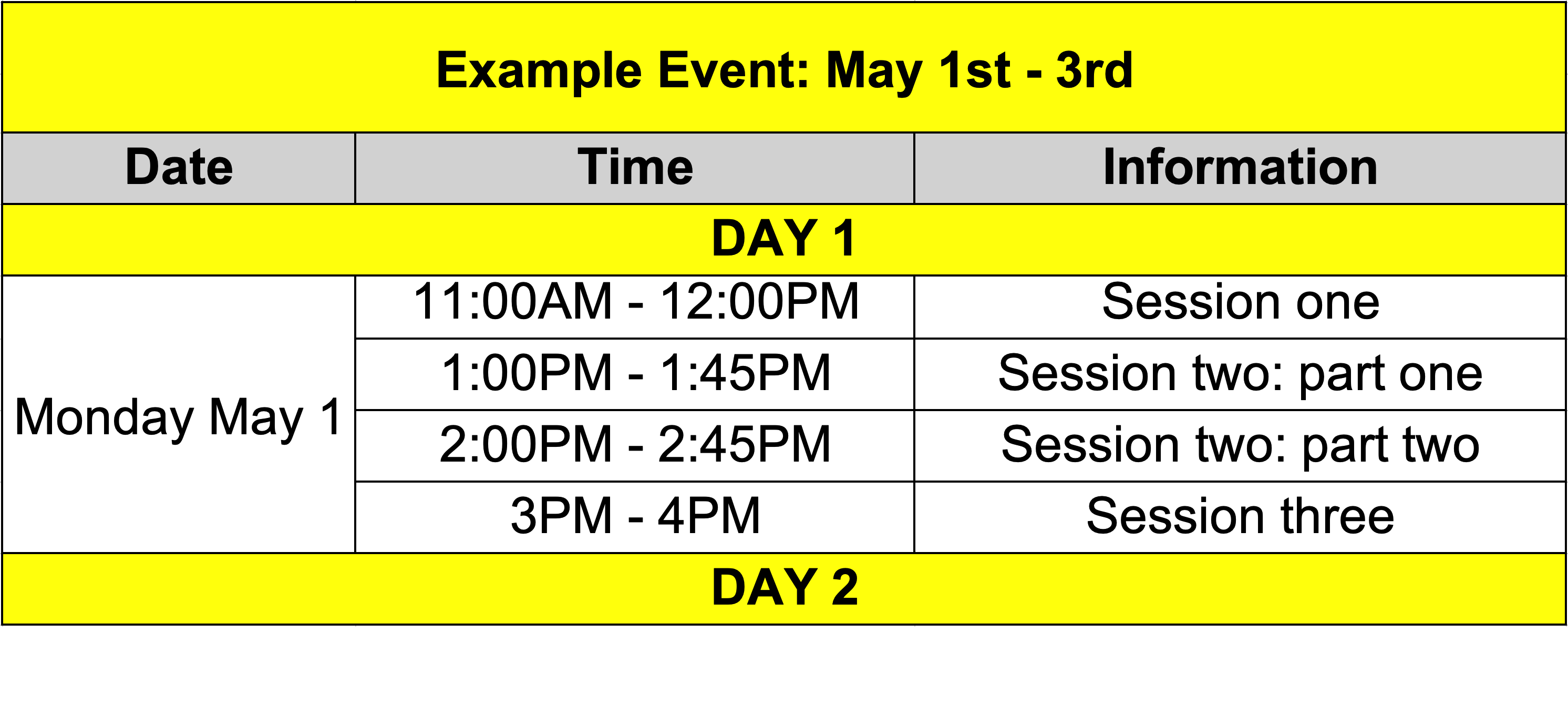}
\caption{Original spreadsheet with merged cells.}
        \label{fig:schedule:nonadapted}
    \end{subfigure}
    \hspace{0.001\textwidth}
    \begin{subfigure}[b]{0.355\textwidth}
        \centering
        \includegraphics[width=\textwidth]{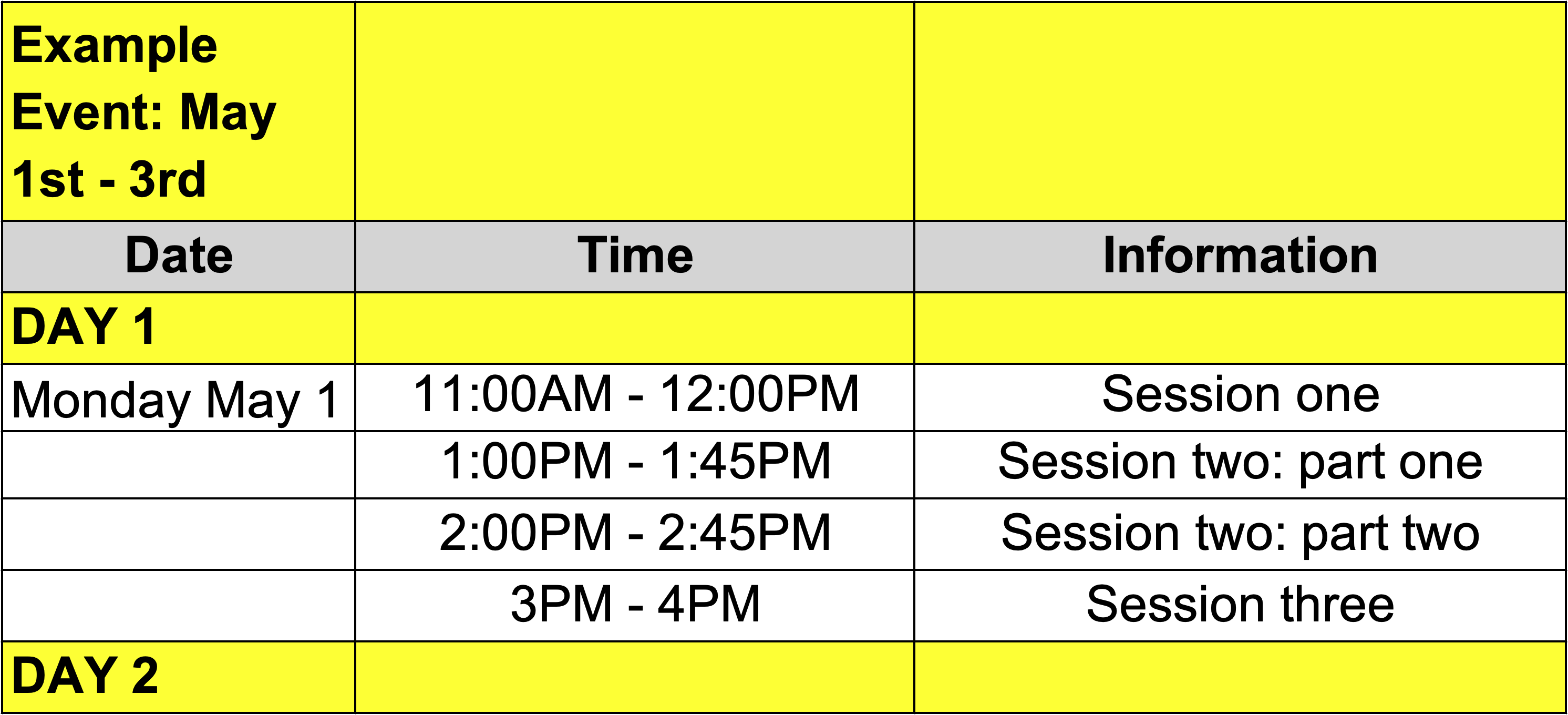}
\caption{P11\sub{LB} unmerged cells.}
        \label{fig:schedule:adapted}
    \end{subfigure}
    \hspace{0.001\textwidth}
    \begin{subfigure}[b]{0.255\textwidth}
        \centering
        \includegraphics[width=\textwidth]{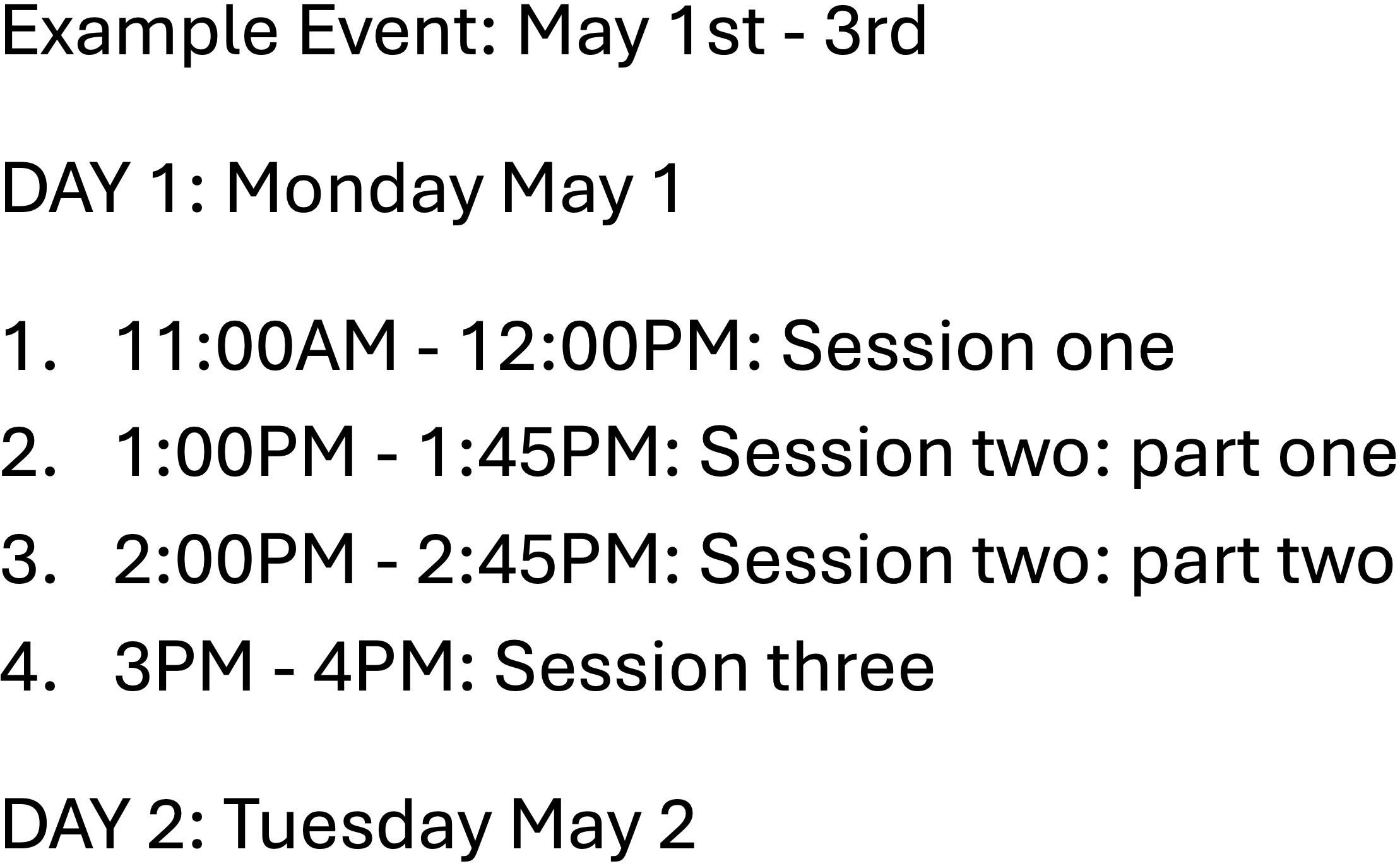}
\caption{P12\sub{LB} suggested a list.}
        \label{fig:schedule:adaptedlist}
    \end{subfigure}
    
    \caption{An example of removing visual structure in spreadsheet. (Confidential content is replaced.) }
    \label{fig:schedule}
    \Description{
The image consists of three subfigures labeled (a), (b), and (c), each demonstrating a different way of formatting an event schedule. (b) and (c) are adaptations of (a).  

(a): Original spreadsheet with merged cells

This shows a spreadsheet where some cells are merged for formatting. The title, “Example Event: May 1st - 3rd,” is merged across the top. Below, the “Date” column contains a merged cell for “Monday May 1,” which spans four rows corresponding to multiple time slots in the adjacent “Time” column. 

The “Time” column lists:
11:00 AM - 12:00 PM
1:00 PM - 1:45 PM
2:00 PM - 2:45 PM
3:00 PM - 4:00 PM

The “Information” column aligns session details with each time slot:
Session one
Session two: part one
Session two: part two
Session three

The merged-cell format allows the “Monday May 1” entry to appear once, visually grouping all related rows. Below this, another section for “Day 2” is visible, also merged to span acrosos multiple columns. 

(b): P11_LB unmerged cells

This panel shows the same spreadsheet, but the cells have been unmerged to provide a different structure. The “Date” column has “Monday May 1” only for the first session, and is empty for other time slots in the “Time” column. The “Time” and “Information” columns remain unchanged. “Day 1” and “Day 2” are unmerged to be only in the first column. 

(c): P12_LB suggested a list

This panel presents the schedule as a list, abandoning the table structure entirely. The title, “Example Event: May 1st - 3rd,” appears at the top, followed by a breakdown of the schedule.

The list begins with “Day 1: Monday May 1” and includes each time slot followed by the session information in text format as a numbered list, for example: 1: 11:00 AM - 12:00 PM: Session one. 

}
\end{figure*}

Representations often rely on visual structure to convey meaning and importance, but visual structure is often not accessible in non-visual ways or poorly rendered by assistive technology. This creates difficulties for BLV colleagues even if the content itself is technically accessible (\textit{n}=10). 
For example, P1\sub{TB} encountered a table in a document with grouped columns, which to him \textit{``[didn't] line up and it's really hard to read from [trying] to use JAWS\footnote{\url{https://www.freedomscientific.com/products/software/jaws}} [screen-reader] to get the information.''} P11\sub{LB} encountered a spreadsheet with grouped columns and \textit{``mistakenly believed that each [header] only had one [column],''} until she accidentally discovered the hidden information by checking cell by cell (e.g., in Figure~\ref{fig:schedule:nonadapted}, the cell ``DAY 1'' spans across multiple columns but this structural information is not represented non-visually). P1\sub{TB}, P8\sub{LB}, P13\sub{TB} and P21\sub{LB} also described how merged cells in tables created confusion, which aligns with Wang et al.'s findings~\cite{Wang_Wang_Jung_Kim_2022}. 
Additionally, P2\sub{S} described documents received from external sources as \textit{``accessible, but messy''} and P21\sub{LB} adds that \textit{``everyone has their own way of organizing [the structure].''} 
Participants who have low or residual vision using screen magnification also struggled due to losing context of the overall visual structure (\textit{n}=8).
P12\sub{LB} noted: \textit{``my level of magnification means that not all of the [visual layout] are visible at one time on my screen, ''} so \textit{``I have to use the lower magnification and lean toward the screen''} which caused physical discomfort. This is also echoed by P28\sub{LB}.

\paragraph{Workaround: Removing Structure} 
In response, it was common for BLV participants to remove visual groupings and prioritize linear and simpler structures for content access as a workaround
(\textit{n}=12). For example, P15\sub{TB} shared that when receiving a spreadsheet, \textit{``sometimes I'll copy it and paste it in a Word document [...] and I can read it better.''} 
P11\sub{LB} removed grouped cells to avoid getting lost (Figure~\ref{fig:schedule:adapted}). 
P3\sub{TB} relied on optical character recognition (OCR) to \textit{``get rid of [the structure]''} entirely to access content as plain text. P16\sub{TB} considered \textit{``brevity''} is important: \textit{``Less is more.''} P12\sub{LB} advocated to \textit{``adjust to thinking more about simpler ways of sharing information [...] Focus on the content rather than the form.''} P7\sub{LB}, P8\sub{LB} and P12\sub{LB} expressed preference for: \textit{``a Word document with a list - It's very simple, very low tech, doesn't look great probably [... but] it may be the thing that people buy into the most''} (P12\sub{LB}, e.g., Figure~\ref{fig:schedule:adaptedlist}). This echoed with P3\sub{TB}'s practice of turning tables into lists, supported by accessibility guidelines~\cite{GovernmentDigitalService_2024, Malone}.

While linearization of representations improved access by removing inaccessible visual structure, 
it also introduced new challenges. It could create friction for sighted colleagues who relied on visual markers (\textit{n}=6), highlighting the bidirectional nature of accessibility challenges. 
P15\sub{TB} recounted sending a long, unformatted document: \textit{``It didn't work out well [... My sighted colleague couldn't] read through pages and pages of text without any formatting: no bold, no headings, nothing visual that could give them some markers.''} 
Participants also mentioned that the resulting linear representation was \textit{``very time-consuming''} to read (P3\sub{TB}), \textit{``never identical''} (P2\sub{S}), making work \textit{``less productive and effective''} (P7\sub{LB}). 
Another drawback of linearizing a representation is the loss of its computational features (\textit{n}=8). 
When P1\sub{TB} P2\sub{S} converted their inaccessible database view into a list of client information in a Word document, this allowed him to read the information, but useful features such as sorting and filtering were not retained. As P1\sub{TB} noted, \textit{“I had to remember to do things in the same order,”} manually replicating functions that could be automated. 
This workaround highlights the compromise BLV workers often face---sacrificing functionality for access, and
\textit{''takes longer time, and accuracy of accessing the information goes down''} (P7\sub{LB}).

\paragraph{Workaround: Adding Accessible Semantics}
Participants utilized accessible structural elements for navigation to supplement the visual structure in representations (\textit{n}=11). 
This is not just a workaround but a solution that improves the representation by creating non-visual access to information structure. 
P14\sub{LV} described a consistent practice of creating an accessible semantic structure \textit{``that include headings, a table of content, and hyperlinks''}, providing a navigational outline, signposts, and shortcuts to content. 
P13\sub{TB} used headings as reference points during collaborative editing: \textit{``[P12\sub{LB}] can tell me verbally [...] which heading [...] to look for.''}
P8\sub{LB} highlighted the benefit: \textit{``If you have the headings in the columns properly labelled [and interactively announced by screen reader ...] you’re not trying to memorize what the column was, what the row was.''} P13\sub{TB} added: \textit{``You know exactly where you are.''} 
Still, P8\sub{LB} and P11\sub{LB} mentioned that not all representations have complete accessible navigational features, hindering their work.

BLV participants also emphasized the role of shortcut keys in navigating semantic structure, particularly when using screen-readers (\textit{n}=5). 
P11\sub{LB} provided an analogy: \textit{``They are our mouse on the keyboard.''} 
The complexity of database interfaces made shortcut keys necessary: \textit{``If you don't have a handle on your shortcut keys, you're going to get really frustrated with that database because trying to navigate through it using your up-and-down arrow keys only is not doable''} (P11\sub{LB}). 
However, P11\sub{LB}, P13\sub{TB}, P17\sub{LB} and P18\sub{TB} mentioned that it takes training to reach a level of proficiency with assistive technology for quick navigation. P13\sub{TB} explained: \textit{``So many, many, many, many, many keystrokes to remember [...] and not everyone does remember all the keys.''} 
P11\sub{LB} advocated for dedicated training, suggesting a \textit{``Shortcut Keys 101''} course.

\subsubsection{System Feedback When Editing}
\label{subsubsec:feedbackEditing}

\begin{figure}[h!]
    \centering
\includegraphics[width=0.4\textwidth]{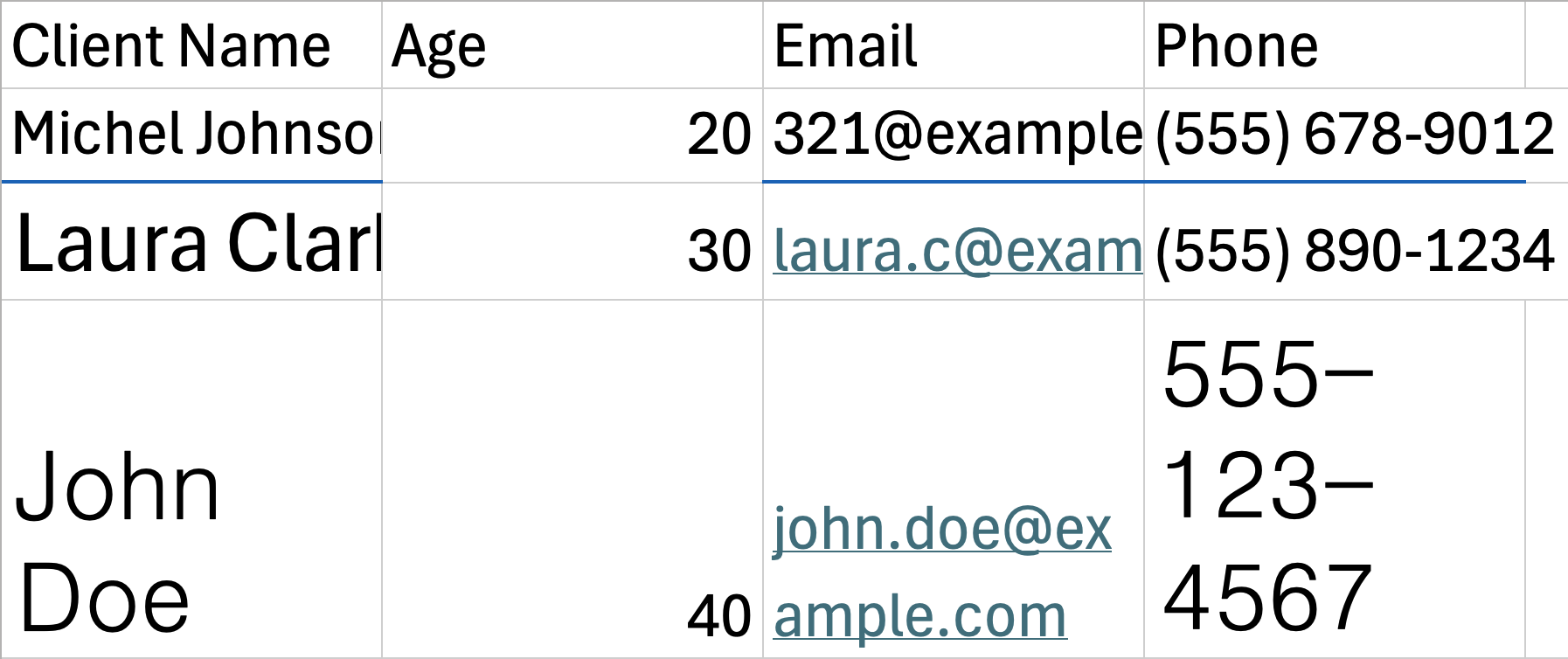}
\caption{P11\sub{LB}'s table created with screen-reader. (Confidential content is replaced.) }
    \label{fig:table}
    \Description{
The table consists of three rows of client data, each containing a name, email address, and phone number, but formatting inconsistencies and errors are present throughout. 

In the first row, the client name is “Michel Johnson,” displayed in a standard font and size the letter 'n' is cut off. The associated email, “321@example,” is incomplete, as the domain portion is missing or cut off. The phone number, “(555) 678-9012,” is correctly formatted and fully visible but spills over the next column. 

In the second row, the client name is “Laura Clark,” which appears in an oversized font and cuts off the last name. The email address, “laura.c@example.com,” is a hyperlink with blue colour and underline also gets cut off. The phone number, “(555) 890-1234,” is also correctly formatted and matches the style used in the first row, but also spills over the next column. 

In the third row, the name and phone are oversized. The client name, “John Doe,” is split across two lines due to text wrapping. The email address, “john.doe@example.com,” is similarly broken into multiple lines, further reducing readability. The phone number, “555-123-4567,” omitted parentheses around the area code and also spans over multiple lines. 
}
\end{figure}

Sometimes, even if the content and structure were accessible, the representation's state was not available to BLV workers to verify whether their actions or modifications were successful or resulted in accidental mistakes (\textit{n}=11). 
P4\sub{S} described receiving documents from P3\sub{TB} with unintentional bullet points. This frustrated P3\sub{TB} who complained that \textit{``I don’t know what I did. My screen reader doesn’t tell me that.''} and described the experience as \textit{``hide and seek [...] I guess it's different if you're reading it [visually], but [...] when you're listening to it [...] it's all in the same tone [...] It's hard to tell what's this and what's that [...] Is this showing up right?”}
P14\sub{LV} pointed out that in PowerPoint, \textit{``if the [...] text box is too small [...] and part of the title gets cut off, our screen-reader doesn’t tell us that.''} 
P10\sub{S} observed a spreadsheet by P11\sub{LB} as using \textit{``various fonts with no level of consistency,''} making the representation less \textit{``clear visually''} (Figure~\ref{fig:table}). 
Additionally, P12\sub{LB} pointed out the difficulty of achieving both accessible and visual formatting: \textit{``If you do the underlining and the bold first and then you do the heading [formatting], it erases the other [semantic] formatting.''} 
P13\sub{TB} echoed: \textit{``They don't make it compatible to both.''} 

\paragraph{Workaround: Manually Tracking Editing Details}
To help address the lack of reliable feedback, BLV workers had to spend significant effort to carefully check editing details (\textit{n}=12). P12\sub{LB} described the process as: \textit{``painstaking[...] keeping track of every heading [...] colon and every semicolon [...] making sure that everything is correct, and it just doesn't seem like the best use of time.''} 
Nonetheless, P3\sub{TB} noted the importance of achieving visually professional outcomes, especially in shared and external-facing representations: \textit{``It’s not for me, but for the other person [...] you want it to look like you know what you're doing. Professional, right?''}
P15\sub{TB} however described sending an official letter without a letterhead because \textit{``I couldn't make it myself''} and \textit{``it probably would look really weird''} if she tried. She acknowledged this was \textit{``not the most professional''} but had to be done for her to complete the task independently.

\paragraph{Workaround: Asking Sighted Colleagues for Verification}
When it was impossible to independently verify editing changes, BLV participants resorted to asking their sighted colleagues for help (\textit{n}=16). P21\sub{LB} was \textit{``unable to properly format information according to the company standards,''}
forcing him to rely on P22\sub{S} and P23\sub{S} to \textit{``[make] it look pretty using the company template''} (P22\sub{S}). 
For P2\sub{S}, documents converted from PDFs to be accessible to P1\sub{TB} were often \textit{``visually horrible''} and she had to \textit{``pretty [them] up''} because \textit{``it's really important that things are in certain spots [...] visually.''} 

However, this reliance on colleagues came with challenges, such as workload conflicts (\textit{``They’re busy, too.''} [P15\sub{TB}]) and cognitive strain (feeling \textit{``drained''} [P10\sub{S}]; \textit{``I definitely have to think harder [...] try and search for the right word''} [P22\sub{S}]).
Some participants expressed concerns about how the need for sighted assistance might impact their professional progression, which we elaborate in Section~\ref{workpriority}. \todo{To fill}

\subsubsection{Cognitive Support}

Representations in participants' work often used visual encodings to relieve cognitive load and support decision-making, such as overviews, aggregations, and comparative displays. 
They cannot be accessed non-visually and fail to provide externalized cognitive support, forcing BLV participants to rely heavily on memorization 
(\textit{n}=11). When understanding a dataset, P7\sub{LB} had to rely on the raw data points, ``putting pieces together in my head based on the numbers that are read aloud,'' rather than analyzing patterns from the visualizations which she could not access. She also mentioned that \textit{``cognitive load is definitely a source of stress and it can be daunting.''} 
Similarly, P21\sub{LB} shared: \textit{``you probably take up more of your [...] brain RAM to do some things that aren't that hard''} and \textit{``memorization is a skill that all blind people need to develop to function well.''}

\paragraph{Workaround: Building Internalized Representations}
Participants described creating mental representations of the information similar to the visuals (\textit{n}=13). P12\sub{LB} explained using \textit{``a spatial landmark - I memorize where things are relative to [it ...] When sharing my screen with someone [sighted ...] I ask them to verbally tell me where the button is relative to a landmark I know.''} 
P8\sub{LB} shared a similar process: \textit{``I need to have it mentally represented in a special way [...] I still try to visualize the data in the format that I think I'm getting from a table.''}  

\paragraph{Workaround: Reducing Information Load}
Participants also acknowledged the need for filtering and summarization (\textit{n}=15) to help understand information when there is a lack of cognitive support. 
P8\sub{LB} shared: ``I extract the information that I need instead of using a big table with a lot of numbers. I break it down into groups [...] and represent them differently in a way that makes more sense to me.''
Alternatively, P1\sub{TB} said: \textit{``because of the volume, a lot of things are represented [...] in [...] a short or summary form to make it easier to go through.''} 
P18\sub{TB} turned to AI tools: % to \textit{``get a summary.''}% : 
\textit{``[I would] plug it into something like Gemini or ChatGPT and get a summary.''} However, P12\sub{LB} expressed reservations: \textit{``I just don't trust that yet.''}

\paragraph{Workaround: Verbal Mediation with Sighted Colleagues}
Participants engaged in back-and-forth verbal interactions as a way to bridge visual and non-visual understanding (\textit{n}=18). P18\sub{TB} stated: \textit{``Verbal communication is huge for us.''}
P21\sub{LB}, P22\sub{S} and P23\sub{S} frequently engaged in verbal discussions using detailed descriptions and targeted \textit{``key questions''} from P21\sub{LB} to bridge the gap between visual and non-visual understanding. 
This sometimes involved trial and error, with P22\sub{S} actively \textit{``gauging [P21\sub{LB}]'s reaction --- If I say something in one style, how quickly does he respond to it? Or is he a little bit confused by what I've been saying?''} P22\sub{S} noted: \textit{``It might be simpler to say fewer things [to start], but at the end [...] we need to dive into the details.''}
P10\sub{S}, P11\sub{LB} and P14\sub{LV} shared examples of presentation coordination, adjusting their delivery based on real-time \textit{``verbal cues.''} 
P1\sub{TB} relied on verbal communication with P2\sub{S} to navigate through documents,  and P15\sub{TB} relied on back-and-forth conversations with P20\sub{S} to discuss requirements and refine designs when creating graphics. 
Several participants also relied on sighted colleagues to filter and prioritize information. P16\sub{TB} would ask a coworker to identify \textit{``important''} sections in lengthy documents. P24\sub{LB} did the same to \textit{``know what's the key point to look at.''}

\subsection{Accessibility As a Work Priority}
\label{workpriority}

Part of the solution to technical challenges is social, where BLV and sighted colleagues collectively work around the representational challenges to enable teamwork.
We found that such social dynamics are influenced by the workplace stigma of overlooking or depriortitizing accessibility in the design and choice of shared information representations that shared by colleagues, clients, or organizations for work. 
Participants emphasized the pervasive ableism embedded in workplace expectations and structures (\textit{n}=10). 
P16\sub{TB} acknowledged:
\textit{``Let's face it: we live in a sighted world, whether we like it or not. And most people learn quite visually [...] Being able to represent things in multiple ways is very important to me.''} 
P7\sub{LB} noted how our conceptions of leadership exclude the experiences of blind leaders: 
\textit{``the leadership role in mainstream society is portrayed in a very narrow way: so you're independent, you do things on your own, and you don't ask for help [... If] you have some kind of disability and you can't do things the same way as a fully sighted individual, you are forced to basically have a different way of doing things and putting yourself forward.''} 
P18\sub{TB} put this eloquently: \textit{``If we had the information that we need to interpret the world around us, we wouldn’t be disabled.''}
In addition, external work constraints also contribute to accessibility challenges. 
P8\sub{LB} explained financial and logistical factors that often hinder accessibility efforts: 
\textit{``Most [designers] do not have the resources planned or available to invest as much as they can in technology [implementation of accessibility ... but they] need to prioritize [... the small] percent of individuals that cannot fully live their lives.''} 
P10\sub{S} also emphasized the systemic issue:
\textit{``A lot of it comes from these outside vendors and when they're designing products they're not really using a universal design method [...] because these companies don't really have to be accessible to get business." }

Participants also described the typical workplace culture as one that prioritizes productivity over accessibility 
(\textit{n}=11). P8\sub{LB} critiqued this tension:
\textit{``The measure of success in an organization is extremely important because if the measure is how you process data, then we're not measuring well. A lot of us are not going to measure [up, when measuring] how fast you are at processing [...] 
I think it's a common need for all of us to be recognized, to be valued, to feel like we're valuable in what we do, in what we bring to the world.''} P12\sub{LB} agreed: \textit{``It's important to measure [...] things like whether or not you want a team that is [...] inclusive in its attitude [...] They're hard to measure, but they're important.''} 
P26\sub{S} explained how competing priorities often lead to overlooking accessibility: 
\textit{``It boils down to [...] workload and competing priorities [...] It also depends on [...] the final audience for things like websites, obviously, you should have a higher focus [on accessibility]. But maybe [not] if it's a one-off project that [...] there's only so few people viewing [...] It sounds so bad to think of it that way, but that's what it boils down to sometimes.''} 
One impact of not prioritizing accessibility in the workplace culture 
is a lack of awareness of information accessibility issues, leading to an attitude of ignorance (\textit{n}=13). 
For example, P28\sub{LB} shared how her colleagues \textit{“just forget”} about others' information access needs.
Moreover, 
BLV employees might not be open to 
disclose their ability needs and request accommodations (\textit{n}=15). P19\sub{S} and P20\sub{S} observed: \textit{``Not everybody wants to share''} (P19\sub{S}) or be \textit{``open about their disability''} (P20\sub{S}).  
Participants are concerned about how the need for assistance might impact their professional progression (\textit{n}=10). P7\sub{LB} shared that this \textit{``do[es] take away from the effectiveness of me being able to portray myself in that leadership role.''}
P21\sub{LB} reflected: \textit{``I think my career path has probably been affected in some ways by my vision loss --- I can't provide the full package [...] I can be outpaced by some sighted colleagues, even though we have very similar skills and abilities.''} 
Furthermore, P17\sub{LB} experienced the emotional toll of their own internalized ableism:
\textit{``I was always embarrassed to ask people [...] I don't reach out and ask people for help because I didn't want them to know how bad I am [...] and how much I struggle [...] I was more hiding my identity [...] I was insecure [about] letting people know that I was blind and or I was losing my eyesight.''}

An accommodating workplace culture is key to shift workplace priorities to consider accessibility, and  top-down commitments from leaders and organization levels are important to \textit{``set the tone culturally''} (P10\sub{S}; \textit{n}=11). P7\sub{LB} explained: \textit{``leading by examples and having intentional conversations''} about \textit{``inclusion''} can establish expectations for accessibility. 
P8\sub{LB} emphasized the importance of social and technical accessibility: \textit{``As a manager [...] we are [aiming to] be diverse, to be accessible, but not just technologically accessible, right? To be accessible to [...] 
different ways of accepting people's work.''}
P11\sub{LB} highlighted that organizational commitment is crucial by implementing accessible representations and providing training resources, and emphasized organizational values supporting \textit{``inclusion, diversity, equity, and accessibility [...] There has to be a precedent from top down.''}

\subsection{Social Adaptations for Representational Accessibility}
\label{socialadaptation}

Beyond organizational approaches to make accessibility a priority in information work, 
P12\sub{LB} noted that bottom-up efforts needed: \textit{``The grassroots [...] level [...] is the one getting the information and actually doing it. And so as long as they're communicating, and the management is allowing [them] to have the autonomy to make it better, then yeah, things do get better over time. And of course, it all hinges on good working relationships.''} 
\textit{``Grassroots''} efforts (P12\sub{LB}), individual \textit{``empathy''} (P7\sub{LB}), and \textit{``trust''} (P14\sub{LV}) can create an open environment from the bottom-up (\textit{n}=14). P3\sub{TB} and P23\sub{S} emphasized the power of \textit{``kindness.''} Shared lived experience also helps foster natural understanding: \textit{``Coming from a marginalized perspective [... has] a big impact on building this culture of inclusion. There's a sense of empathy''} (P7\sub{LB}). This open environment can facilitate smoother information exchanges, particularly when collective efforts are needed to make representations accessible. Below, we elaborate on the four types of processes that progressively enable team members to build awareness and work towards interdependence in information work.

\paragraph{Advocating for Representational Accessibility}

Given the lack of awareness of representational accessibility, 
the responsibility of communicating accessibility requirements often falls on the BLV employees.
Participants noted that representational modifications to improve access are sometimes made only after explicit requests from BLV colleagues (\textit{n}=11).
An email exchange between P3\sub{TB} and an external service provider demonstrated the default assumption that everyone can access information in the same visual way. P3\sub{TB} informed: \textit{``PowerPoint is similar to PDF; Word works best.''} 
P4\sub{S} and P6\sub{S} adapted their practices after a colleague's feedback \textit{``that my texts weren't bold and I started making it bold on the meeting minutes"} (P6\sub{S}). \textit{``[P5\sub{LV}] said: Can you do [enlarged fonts]?''} (P4\sub{S}). 
P8\sub{LB} shared similar experiences: 
\textit{``They asked [... to] `send me the link that I can look at it on my own time.' So that's what we created. Now we have a link and they can check the document whenever they want.''}

While requesting adaptations demonstrates self-advocacy, 
it still relies on other stakeholders' willingness to accommodate. P16\sub{TB} shared that accommodating workplaces
\textit{``are going to understand that and they're going to give you the time [... not having the] attitude of --- Well, if you can't do it my way, then hit the highway.''} 
P19\sub{S} mentioned biased beliefs:
\textit{``Employers think - Oh, we're gonna have to provide all this technology and all the training. And [...] it's gonna be very costly''} but most necessary accommodations for them would cost nothing and other accommodations are typically under \$500~\cite{BeMyEyes}.

Some participants advocate for systemic change and raising awareness about accessibility within their workplaces (P4\sub{S}, P16\sub{TB}) and broader communities (P4\sub{S}, P5\sub{LV}, P17\sub{LB}, P19\sub{S}). 
P4\sub{S} was actively 
\textit{``advocating for accessibility [...] With the [...] contract [documents received,] we'll point out [...] how things aren't accessible, how they could be accessible.''} 
P16\sub{TB} shared examples of how his advocacy efforts \textit{``made the company I was working for a lot more aware of using an accessible database.''} Four participants advocated for accessibility as a fundamental right, moving beyond the idea of accommodations as favours.

\paragraph{Assuming Representational Access Needs}

Individuals who have a basic level of awareness might assume what accommodations are needed without consulting or knowing the recipients of the information. Such assumptions, though often well-intentioned, can reflect limited understanding and result in inappropriate or ineffective accommodations.
(\textit{n}=8). 
P9\sub{S} initially believed non-adapted document formats were sufficient for her BLV colleague: \textit{``She hasn't said anything to me so I think it's okay.''} P9\sub{S} later reflected: \textit{``Whenever I have to send a document in the future for anyone, it's always going to be at a certain font size [with high contrast.]''}
P6\sub{S}'s assumption about screen-reader functionality revealed a knowledge gap: \textit{``I'm not really sure exactly how the screen-readers [work ...] If it's not at the top of their email list, then they might not be able to scroll down.''} 
This passive approach risks overlooking individual preferences and their need to achieve efficiency.

Providing multiple formats is encouraged as an accessibility practice (\textit{n}=12), as P6\sub{S} advised: \textit{``I think you've got to take information and represent it in multiple means. I don't think there's a one-size-fits-all.''} 
P3\sub{TB} described such an instance leading to increased access: \textit{``[My collaborator] always sends me a second email with what she can in Word documents as opposed to the previous PDF ones.''} 
P13\sub{TB} suggested such practices should be standardized: \textit{``There should be a guideline if you're going to be sharing a document to everyone who's going to be accessing it.''} 
However, participants also highlighted possible assumptions in these practices: \textit{``The assumption that for people who are visually impaired, that bigger is better, which isn't always the case.''} (P29\sub{LV}) and \textit{``if we zoom in too far then we can't read it''} (P28\sub{LB}) because they cannot utilize \textit{``their peripheral [vision] to see more of the sentence.''} (P30\sub{LB}).

\paragraph{Inquiring about Representational Access Needs}

Assumptions about what works best might not be universally applicable, highlighting the importance of seeking individual feedback and making adjustments accordingly.
Some participants actively sought to understand the specific ability needs and preferences of information recipients to provide accommodations, lightening the burden of advocating for accessibility (\textit{n}=9).
For example, P7\sub{LB} directly asked her colleague: \textit{``What is the best way for me to send this information to you? Or what can I do to communicate with you or continue to communicate with you effectively [if your eye] sight is changing?''} P27\sub{S} suggested a survey which would \textit{``[ask] how they could make their presentation more accessible.''}
Participants also expressed a genuine desire to understand and adapt. 
P9\sub{S} conveyed a desire to learn from BLV colleagues' experiences: \textit{``I would love to [...] discuss more about their own experiences. I think it's important because then I have a better understanding of how to interact with [them].''} She also added: \textit{``I also probably should download JAWS so that I can have a better understanding.''}

\paragraph{Building Interdependence for Representational Access}

Participants who are fully aware of the accessibility needs of their colleagues proactively ensure that information representations are accessible appropriately. 
This goes further because actions to adapt and understand individual abilities and representational practices become an ingrained part of the process of work itself, not requiring additional attention or questioning. Participants emphasized the need to understand and respect individual accessibility needs (\textit{n}=11). 
P7\sub{LB} actively adapted her representations of information to suit individual preferences: \textit{``Depend[ing] on the audience [...] I do modify how I receive information or how information is shared [...] So that I actually can respond appropriately.''} P23\sub{S} demonstrated a deep understanding of P21\sub{LB}’s preferences:
\textit{``You kind of know what they're capable of, what they like doing, what they don't like doing, and [...] how they prefer things.''} 
P8\sub{LB} also emphasized task assignments considering individual abilities: \textit{``I hire people for those specific skills [...] so their abilities matter for the tasks that I'm giving them [...] to delegate that specific task to somebody that has more ability to do it [...] I won't give something [that they are] going to struggle with.''}

Participants shared that becoming aware is an ongoing learning process (\textit{n}=11). P2\sub{S} captured the dynamic of \textit{``trial and error --- We just work it out.''}
P20\sub{S} described his own realization of the need for image descriptions: \textit{``As we were going along, it kind of hit me that [...] regardless of the case, I really should be giving a description.''} 
P21\sub{LB} highlighted contributing factors such as \textit{``exposure to [inaccessible situations, ...] seeing how others work with that person, and just building rapport [...]
It's really just about being human and talking to people and [...] trying to work together.''}

Participants highlighted the balance between independence and assistance to achieve interdependence (\textit{n}=6). 
P4\sub{S} reflected: \textit{``There's a fine balance between doing what is needed or ask[ing] to make things more accessible or easier for someone [...] and just going ahead and doing everything [...]
I could just do it [for my BLV colleague,] and then she's like: `Why don't you let me?'''} 
P25\sub{S} explicitly identified the trade-off: 
\textit{``Keeping in mind that independence is also important [...] It's also that trade-off of being helpful to the degree that anybody receiving help would want.''}
P16\sub{TB} elevated the importance of interdependence: \textit{``As a community, if we could all be interdependent and rely on each other and also support one another, then I think that builds resilience just by virtue of being connected.''}  
\section{Discussion}
The previous section reported the key evidence and phenomena exposed by our data, organized in three main themes: how representations fail to support mixed-ability team work, how the institutional and group rules and culture of work affect the choice and effectiveness of representations and how teams progressively might engage in increased awareness of the representational needs of each other to achieve a more desirable mixed-ability workplace. In this discussion, we first interpret the significance of each of these three sets of results, relating the findings to the growing corpus of knowledge on mixed-ability work teams, and then explain how the three themes relate to each other. The higher-level overview allows us to, additionally, offer indications for future evidence-compatible approaches to address some of the challenges.

\subsection{Representational Failures and Their Implications}
In Section~\ref{repproblems} we identified four types of challenges (and some solutions) related to the accessibility and choice of representations. 
The first type is straightforward: if information is not representable for some team members, collaboration becomes impossible. This remains as a foundational barrier to collaboration, even though this is often covered by accessibility literature and standards (e.g., ~\cite{chartAccessReview, UI/UXGuidelines}. 

Slightly less obvious is the role of information structure (rather than just content) in acquiring and navigating information. It is visually clear for sighted users but is inaccessible non-visually unless explicitly encoded semantically. 
For example, the structure in word processors is rendered as visually separated hierarchies of titles with larger or bold font. This is invisible to BLV workers without semantically distinguishing the titles (as heading objects). 
Interestingly, the opposite also happens; BLV workers sometimes produce documents that lack sufficient visual structure, making them impractical for sighted co-workers.
These mismatches demonstrate that accessibility failures often arise not from missing information, but from incompatible structural conventions. Our findings remind us that structure and navigation cannot be ignored, and it is compatible with previous work~\cite{Wang_TabularDataBLV}.

In a similar way, the evidence also reminds us that we need to consider editing and creating shared representations in working environments where BLV workers need to be able to contribute. 
Unfortunately, typical editing and creation functionalities in interfaces tend to be difficult to make accessible non-visually due to the much more complex sets of actions, for which feedback and feedthrough are also fundamental. Although there recent research have explored creating and editing a wider variety of document types by non-visual means (e.g.,~\cite{Zhang_Wobbrock_2023,Zong_PedrazaPineros_Chen_Hajas_Satyanarayan_2024}), and GenAI approaches has made it possible to generate almost anything from text (e.g.,~\cite{Lee_Kohga_Landau_O’Modhrain_Subramonyam_2024}), the reality for most BLV workers in mixed-ability teams is that only interfaces based in text and efficiently supported by screen reader functionality practically allow editing. 
Even in applications that have long supported accessibility for BLV users (e.g.,~\cite{Lee_Zhang_Herskovitz_Seo_Guo_2022}), the consequences of inadvertently modifying the document in a visual way can be mortifying for the BLV worker (who might suffer a perceived or real loss of reputation) and also impede the work of the sighted collaborator (Section~\ref{subsubsec:feedbackEditing}).

A final set of challenges brought up by participants relates to the additional cognitive load that they have to invest to understand and make sense of information that is technically accessible, but not efficiently. Visual representations such as charts, tables and diagrams have continually evolved over many decades and offer substantial cognitive advantages (e.g., quick overview~\cite{BenShneidermanMantra,TableJiPerinNacenta}, emergent insights~\cite{larkinWhyDiagramSometimes1987}). However, we do not seem equally committed to providing non-visual interfaces that are similarly efficient. Perhaps it is much harder, but we think that the research community has simply not yet invested sufficient time and effort to achieve comparable results.

Taken together, we think that our characterization of representational challenges is useful to surface recurring pitfalls and that the (bottom-up) representational approach that we use to examine work problems is fruitful in exposing them. In fact, the four dimensions resemble the WCAG guidelines~\cite{WCAG} and the ladder of diagram accessibility proposed in~\cite{Zhao_Nacenta_Sukhai_Somanath_2024}, with related issues validated in prior work (e.g.,~\cite{HamideKerdar_Bächler_Kirchhoff_2024, Jordan_VanHyning_Jones_BradleyMontgomery_Bottner_Tansil_2024}).

\subsection{Institutional and Cultural Rules and Expectations}

Often in a very different tone to their discussion of specific representations, our participants emphasized how institutional values, their leadership, or organizational culture determine their ability to meet performance expectations. 
While our diary study provided temporal snapshots of such social structures in a workplace, our follow-up interviews allowed participants to reflect on how these daily barriers can result from the broader systemic issues over a longer time (sometimes years). 
This reflects an awareness of how top-down solutions such as standards, rules and initiatives can help address the representational challenges. This supports Marathe and Piper's argument about the paradox in which technology companies claim to value accessibility yet their BLV employees still encounter barriers from the organizational infrastructure in practice~\cite{Marathe_Piper_2025}. 

The social model of disability~\cite{Shakespeare_2006, Shakespeare_1997} posits that disability is produced by social organization rather than individual impairment. In our data, this manifests as institutional rules amplifying disablement. For instance, deliverable-focused metrics can exacerbate the cost of non-visual work (e.g., in Section 4.2, P8\sub{LV} critiqued the institutional \textit{``measure of success''} in terms of speed, which might disadvantage BLV workers).
While top-down approaches in institutions to address accessibility (e.g., DEI/EDI charters and goals) or even from the government level (e.g., Accessibility Standards Canada\footnote{https://accessible.canada.ca/} and the Americans with Disabilities Act (ADA) Standards for Accessible Design\footnote{https://www.ada.gov/law-and-regs/design-standards/}) 
% It would be difficult to argue that top-down approaches in institutions (e.g., DEI/EDI charters and goals), or even from the government level (e.g., Accessibility Standards Canada\footnote{https://accessible.canada.ca/} and the Americans with Disabilities Act (ADA) Standards for Accessible Design\footnote{https://www.ada.gov/law-and-regs/design-standards/}) have been completely ineffective. Top-down pressure to address accessibility 
are probably useful, these are also likely insufficient to achieve what Mia Mingus calls \emph{access intimacy}~\cite{Mingus_2011}. 
Our data indicates that coming closer to access intimacy requires a relational understanding that goes beyond compliance, emerging through specific interactions between team members rather than broader policy shifts (e.g., in Section 4.1.4, P22\sub{S} and P21\sub{LB} actively adapting their descriptive style to bridge non-visual understanding). It is difficult to argue that top-down approaches are ineffective; they provide a necessary baseline. Yet, without the complementary bottom-up process of forming an appropriate work culture, standards might not achieve the desired outcomes. 
The question is, then, how does the complementary bottom-up process of forming an appropriate culture of work develop? How does it emerge? Our finding points at awareness.

\subsection{The Mechanics of Awareness}
The representational lens that we chose for this study is inherently low-level because it looks at the use of specific representations for specific work tasks in specific circumstances. We think that this bottom-up approach, as well as the diversity of participating teams, enabled us to observe some of the processes that allow mixed-ability teams to work better and achieve something closer to \emph{access intimacy}. This is what we described in Section~\ref{socialadaptation}. People start at an undersirable stage where sighted people's lack of understanding of BLV co-workers representational needs can result in unnecessary work, misunderstandings, and inefficiencies. This might then trigger situations in which workers ask their colleagues to adopt new representations, or they advocate for themselves, 
which may become embedded in workflows or even formal recommendations over time.
However, these might still be suboptimal despite the goodwill of everyone involved, since certain adaptations might be unnecessary or even detrimental for certain people, and it is often difficult for people to adapt their work habits if they do not sufficiently understand what is required by their co-workers. 
Our findings suggest that it is mostly through personal interactions, often one-to-one, that people can develop effective workflows by choosing the appropriate shared representations and using them with their colleagues in mind. This is consistent with a relational view of disability~\cite{Shakespeare_2006, Thomas01012004, Garland-Thomson_2014}, which suggests that ability and disablement are negotiated in social relations rather than fixed states. The specific variants of workflows and what triggers them are studied in detail in~\cite{Zhao_Nacenta_Sukhai_Somanath_2026_CHI}. Incidentally, the way in which people learn to adapt to each other's representational needs also supports building up trust, group cohesion and efficiency. Many participants showed awareness of the importance of the process and willingly participated in it.

Importantly, there is no guarantee that everyone will follow through with this evolving process. It is possible that the environment, the rules, the incentives, or the configuration of different groups prevent progression in different ways. We suspect that more transactional environments (e.g., based exclusively on deliverables and performance measurements and exemplified in the data by client relationships mediated mostly through deliverables) might end up being detrimental in comparison to the relational environments that many of our participants yearn for.

\subsection{Better Technology, Better Rules, or More Awareness?}
One might be tempted to consider each of the three areas from our analysis as suitable areas for the removal of challenges separately. However, we suspect that this is not the best approach because in most cases, representational issues, top-down approaches and the building up of awareness were tightly connected. For example, when P3\sub{TB} expressed frustration with a PDF document that impeded her work (a representational failure), she implicitly refers to her difficulty measuring up to the expectation of efficiency (rules and expectations). Eventually P4\sub{S} requested the owner of the document to change it to Word to enable P3\sub{TB}'s work, and advocated for a wider change of workflows and choice of representations to enhance awareness of such issues. This shows how representational problems are embedded within organizational structure and expectations, and addressed through both technical fixes and social processes that build awareness among colleagues. 

Although individual technical improvements in representations might still make a difference, we advocate for a more holistic approach in which designers of representations and workflows consider also organizational expectations and rules as well as how people build awareness of their representational needs with each other. This recommendation might be difficult to heed, but it is borne from the evidence collected from actual mixed-ability teams that carry out work every day. 
We propose that organizational guidelines should focus on shifting workplace culture to value accessibility as a core component of knowledge work productivity (e.g., in Section 4.2, P8\sub{LV}'s advocacy for companies being open to differences in how people work), rather than on accommodation. 

\begin{figure}[h!]
    \centering
    \begin{subfigure}[b]{0.3\textwidth}
        \centering
        \includegraphics[width=\textwidth]{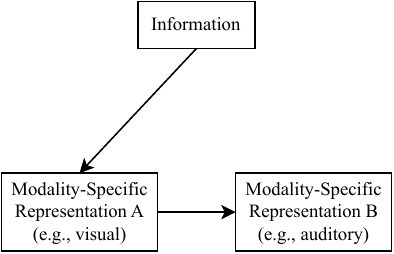}
        \vspace{-0.01\textwidth}
        \caption{Information gets represented visually first, then translated into other modalities.}
        \label{fig:translate}
    \end{subfigure}

    \vspace{0.04\textwidth}
    
    \begin{subfigure}[b]{0.3\textwidth}
        \centering
        \includegraphics[width=\textwidth]{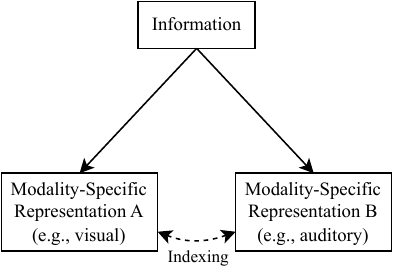}
        \vspace{-0.05\textwidth}
        \caption{Information gets directly represented in multiple modalities.\\}
        \label{fig:direct}
    \end{subfigure}

    \caption{Two ways of creating accessible representations. }
    \label{fig:modRep}
    \Description{
The diagram presents two distinct approaches to creating accessible representations of information across multiple modalities. In subfigure (a), information is first represented visually and then translated into other modalities, such as auditory. This sequential process begins with a visual representation as the primary mode, followed by a secondary adaptation into another modality. 

In (b), information is directly represented in multiple modalities simultaneously. Rather than being translated from a visual source, the information is immediately represented both visually and auditorily in parallel. The dashed arrow between these two modality-specific representations indicates an indexing or synchronization mechanism. 
}
\end{figure}

Nonetheless, digital technology can offer opportunities to address some of the challenges at the representational level, just not in the obvious ways. In his monograph on Representation, Inclusion and Innovation~\cite{lewis2017representation}, Lewis advocates for building representational systems where the information is first represented modality-free, and only then rendered for multiple specific modalities as needed (see Figure~\ref{fig:direct}), rather than the most common current approach, which is to translate visual representations into other modalities (Figure~\ref{fig:translate}). We think that this is a feasible approach to address many of the problems that we observed, where different people require different representations depending on their sensory abilities (e.g., the visual table in Figures~\ref{fig:schedule:nonadapted} and the list in~\ref{fig:schedule:adaptedlist} can co-exist and be cross-indexed to ensure edits made in one update the other). 
% A representative example of this approach is the content-format separation of early web pages, which could easily be rendered and edited independently of their visual properties.

We think GenAI technology can help realize this idea, provided they are thoughtfully embedded within broader organizational frameworks rather than standalone deployment. Specifically,  
GenAI could help in two key ways. First, AI models may enable the reliable automatic extraction of information from existing visual materials to modality-free representations (Figure~\ref{fig:translate}: reversed arrow direction from visual representation back to information; e.g.,~\cite{altGen}). Second, GenAI may support the design and creation of modality-dependent interfaces (Figure~\ref{fig:direct}) that achieve the functional goals that we derived from our study in Sections~\ref{repproblems}: they represent all the information, its structure, enable efficient editing through appropriate feedback and feedthrough, and they support cognition. This is currently a tall order because we have observed that, as the complexity of the representation and its functionality increases (e.g., rich-feature tools, editing), its accessibility is less likely to be addressed appropriately, forcing teams to settle for less capable representations~\cite{Zhao_Nacenta_Sukhai_Somanath_2026_CHI}. Probably the most interesting challenge in this direction is finding ways to support cross-referencing the information (Figure~\ref{fig:direct}: dashed arrow of indexing) from multiple adapted representations that are specific to the modality and abilities of different people in the team. This, while preserving the ability of different participants to become aware of each other's needs, as we advocate in the first paragraph of this section.

As GenAI tools improve and their use becomes pervasive, we should stay vigilant that the promised increase in power does not result in impoverished work environments and fewer opportunities to relate and understand each others' needs.

\subsection{Limitations}

We initially expected more prevalence of using GenAI technology in participants' work, but they reported limited use of it, largely due to privacy and confidentiality concerns. Future work is needed to examine settings where AI use is more established. 

Although we tried our best to recruit teams with different professional backgrounds and participants of different knowledge work roles, the representativeness of the teams sample in this kind of qualitative study is limited. This is partly due to the already substantial commitment required from diary study participants. Although a longer data collection period might capture more varieties of diary data, we believe the five-day period of the diary study is a reasonable balance between data richness and participation burden. 
Regarding the methodology, the diary with interview and the focus group data differ in their methodological affordances, which may affect data comparability. Upon examining the codebook, both data sources share the same codes, with focus group codes being a subset of diary codes. This is to be expected and likely reflects the greater richness and granularity of the diary data. 

Furthermore, our diary study is primarily representation-centric. While representations are foundational to information work, future work should connect our findings to broader knowledge work issues, such as unique challenges of remote or hybrid work, which may amplify barriers for mixed-visual ability teams.

\section{Conclusion}

Investigating accessibility of representations, we confirm that
many barriers persist in today’s knowledge work settings, often requiring costly manual workarounds while creating a sense of exclusion. Although there have been advancements in specific domains of work and many groups are working on translating or creating new information representations in different modalities, the broader experiences of mixed-visual ability teams remained largely unchanged~\cite{Branham_Kane_2015}. Our work investigates, with a representational lens, how teams use external representations, how they can be enabling or disabling, and how teams get around the barriers. 
Our results highlight the importance of interdependence and open cultures of adaptability and inclusivity, which can be strategically leveraged in future designs of representational ecosystems. Based on this perspective on representation accessibility, we also encourage designers to create sets of representations native to different modalities for different team members, as well as leverage GenAI technology, both of which have the potential to empower individuals and mixed-visual teams. 

While knowledge work tools tend to support productivity goals (e.g.,~\cite{google_marketplace_productivity}), we advocate for a broader design scope that accounts for the invisible work required for accessibility. We argue that true workplace flourishing cannot be achieved 
by excluding diverse abilities; instead, we should transition toward ecosystems that support interdependence over individual output: the future of work is not just more productive, but more equitable for everyone.

% \section{Acknowledgement}
\begin{acks}

ChatGPT and Gemini with web access were used to assist in improving language in the authors' existing text and literature search. The paper remains an accurate representation of the authors’ underlying work and novel intellectual contributions.

This research is supported by the University of Victoria, NSERC Canada Graduate Research Scholarship---Doctoral 588775-2024, and NSERC DG 2020-04401.
 
\end{acks}

\bibliographystyle{ACM-Reference-Format}
\bibliography{0-main}

\appendix

\end{document}